# Analysis of Bayesian Classification based Approaches for Android Malware Detection

Suleiman Y. Yerima, Sakir Sezer, Gavin McWilliams

Postprint

# Analysis of Bayesian Classification based Approaches for Android Malware Detection

Suleiman Y. Yerima, Sakir Sezer, Gavin McWilliams
Centre for Secure Information Technologies (CSIT)
Queen's University, Belfast,
Northern Ireland
Email: s.yerima@qub.ac.uk

*Abstract*— Mobile malware has been growing in scale and complexity spurred by the unabated uptake of smartphones worldwide. Android is fast becoming the most popular mobile platform resulting in sharp increase in malware targeting the platform. Additionally, Android malware is evolving rapidly to evade detection by traditional signature-based scanning. Despite current detection measures in place, timely discovery of new malware is still a critical issue. This calls for novel approaches to mitigate the growing threat of zero-day Android malware. Hence, in this paper we develop and analyze proactive Machine Learning approaches based on Bayesian classification aimed at uncovering unknown Android malware via static analysis. The study, which is based on a large malware sample set of majority of the existing families, demonstrates detection capabilities with high accuracy. Empirical results and comparative analysis are presented offering useful insight towards development of effective static-analytic Bayesian classification based solutions for detecting unknown Android malware.

## 1. INTRODUCTION

The Android mobile platform is increasing in popularity surpassing rivals like iOS, Blackberry, Symbian and Windows mobile. The apps available on the Google Play Android market alone are well over 675,000, with an estimated 25 billion downloads (as at October 2012) [1]. At the same time, malware targeting the Android platform has risen sharply over the last two years. According to a report from Fortinet (November 2011), approximately 2000 Android malware samples belonging to 80 different families had been discovered [2]. Since the discovery of the first Android malware in August 2010, more sophisticated families capable of evading traditional signature-based detection are emerging [3].

In February 2012, Google introduced Bouncer to its official app marketplace to screen submitted apps for malicious behavior. No doubt a welcome development towards curbing malware, this has not completely eliminated the problem. Bouncer is based on run-time dynamic behavioral analysis; and possible means of circumventing its analysis process have been demonstrated by Oberheide and Miller [4]. Moreover, other than Google Play, users commonly download apps from third party sources not protected by Bouncer.

According to security experts, the difficulties in spotting malicious mobile apps results in most Android malware remaining unnoticed for up to 3 months before being discovered [2]. Furthermore, Oberheide et al. [5] observed that it took on average 48 days for a signature-based antivirus engine to become capable of detecting new threats.

Clearly, there is a need for improved detection capabilities to overcome the aforementioned challenges and mitigate the impact of evolving Android malware. Hence, in this paper we present Bayesian classification based machine learning approaches that utilize static analysis to enable proactive Android malware detection. The methods are effective in detecting known families as well as unknown malware with reasonably high accuracy. Thus, it is definitely useful in overcoming the limitations of traditional signature-based scanning as well as viable for filtering apps for further analysis by complementary methods or manual reverse engineering analysis by security analysts, thus reducing the costs and effort involved in uncovering new malware samples.

In this paper, three Bayesian classification based approaches for detecting Android malware are presented and analyzed. These are developed from application characteristics obtained through automated static analysis using a large scale malware sample library of 49 known Android families and a wide variety of benign apps. We discuss three viable Bayesian classification models that can be built from statically mining a large collection of apps, and provide empirical results that offer useful insight towards development of effective automated static analysis based solutions for detecting unknown Android malware.

The rest of the paper is organized as follows: related work is discussed followed by the automated reverse engineering and static analysis that underpins the proposed Bayesian approaches. Next, the Bayesian models' formulation is presented. Experiments, results and analyses follow; the paper is then concluded further work outlined.

## 2. RELATED WORK

In the current literature, related work on behavioral based mobile malware detection such as [7], [8] or on-device anomaly detection [9] can be found. Different from the aforementioned, this paper proposes and analyzes off-device, data mining approaches that employ static analysis of Android application packages, whilst avoiding performance bottleneck issues of on-device approaches. Static analysis has the advantage of being undetectable, as obviously malware cannot modify its behavior during analysis [2]. Thus, it has been applied to Android vulnerability assessment, profiling, threat detection etc. For example, ComDroid [10] is a static analysis tool for detecting application communication vulnerabilities. DroidChecker [11] is a tool for detecting capability leakage in Android applications. ProfileDroid [12] is a monitoring and profiling system for characterizing Android app behaviors at multiple layers: static, user, OS and network. RiskRanker [6] provides not only profiling but also automated risk assessment to police Android markets and aid zero-day malware detection. RiskRanker employs a two-order risk analysis system and classifies apps as high, medium or low risk. Profiling and reporting function for Android applications based on static analysis is also presented in [13]. Though the method used in [13] is designed to identify security and privacy threats, unlike the study in this paper, it is not based on data mining or machine learning.

Other existing works that employ static analysis for detection of malicious activities like SCANDAL [14], AndroidLeaks [15], and the framework presented in [16], focus on privacy information leakage. Whereas, the malicious activities targeted by our work extends beyond privacy information loss.

In [17] Blasing et al. presented an Android Application Sandbox (AAS) that uses both static and dynamic analyses on Android applications to automatically detect suspicious applications. For the static analysis part, the code is decompiled and 5 different types of patterns are matched namely: JNI usage, reflection, spawning child processes, services and IPC usage, and runtime requested permissions. Compared to AAS, our methods cover a much wider range of pattern attributes extracted not only from the application code logic but also scrutiny of resources, assets, and executable libraries where malicious payload could be lurking. Additionally, these attributes contribute to ranked feature sets which drive our Bayesian classification models.

In [2] Apvrille and Strazzere employ a heuristics approach based on static analysis for Android malware detection. Their heuristic engine uses 39 different flags weighted based on statistics computed from techniques commonly employed by malware authors in their code. The engine then outputs a risk score to highlight the most likely malicious sample. Our approach shares similarity in the reverse engineering technique, but differs by utilizing Bayesian classification methods that are more flexible and easier to maintain. For example, models can be re-trained as new malware samples are discovered, while features sets can be automatically updated.

In [18], Schmidt et al. employ static analysis on executables to extract their function calls using the readelf command. They then compare these function call lists with those from Linux malware executables in order to classify the executables using learning algorithms. In contrast, our static analysis approach is based on automated analyses of Android packages. Moreover, Android malware samples across a wide range of existing families are employed in our work rather than Linux malware executables.

Other earlier non-Android based papers have explored data mining and machine learning techniques for malware identification including for example [19], [20] and [28]. The authors of [19] apply machine learning methods on a data set of malicious executables where a set of Windows and MS-DOS format executables are utilized while comparing three learning algorithms with signature based detection. While [20] is based on application of data mining methods and SVM to distinguish between benign executables and virus by statically extracting dynamic link libraries and application programming interfaces.

For the Android platform, a paper by Sahs and Khan [21] presented a machine learning approach for Android malware detection based on SVM. A single-class SVM model derived from benign samples alone is used. Contrary to their approach, our classification models are trained with both a wide variety of benign apps and a range of samples from across 49 malware families discovered in the wild. Also, in [22], PUMA (Permission usage to detect malware in Android) detects malicious Android applications through machine-learning techniques by analyzing the extracted permissions from the application itself. Our work leverages not only permissions, but also other code-based properties through automated reverse engineering to investigate our data-mining approach for malware detection. Moreover, our study was undertaken with a larger malware sample set. Different from [22], this paper also provides insight into permissions usage from a different perspective; i.e. in-depth comparative analysis with the use of other viable application properties to underpin the machine learning detection approach.

In summary, the main contributions of this paper different from existing related works in the literature are as follows:

• Novel approaches that apply automated static analysis based Bayesian classification for proactive Android malware detection.

• Extensive empirical evaluation and comparative analysis of the Bayesian classification methods with a large malware sample set from across 49 malware families in the wild.

Our approach for discovery of unknown malicious applications is motivated by the need to bolster existing methods given their limitations. We also note that the significant delay between malware release and eventual discovery is still a critical.

### 3. ANDROID APP REVERSE ENGINEERING

Android applications are written in Java and compiled by the Android SDK tools —along with any data and resource files—into an Android package (APK), an archive file with an .apk suffix. All the code in a single .apk file is considered to be one application and it is this file that Android-powered devices use to install the application. The applications are distributed as self-contained packages that are compressed (ZIP) bundle of files typically consisting of: AndroidManifest.xml (Manifest file), classes.dex (A single file which holds the complete bytecode to be interpreted by Dalvik VM). Other binary or XML-based resources required by the application to run may be held in res/ and assets/ folders.

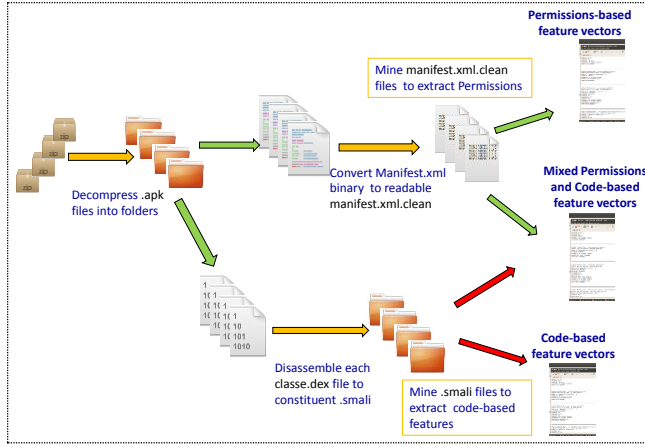

**Figure 1. Automated Android app reverse engineering and data mining for Bayesian model(s) feature extraction with the Java-based custom built APK analyzer.**

The Android application is built from four different types of components: Activities, Services, Broadcast Receivers, and Content Providers [23]. An application must declare its components in a Manifest file which must be at the root of the application project directory. Before the Android system can start an application component, the system must know that the component exists by reading this file. The Manifest file also states the user permissions that the application requires, such as internet access or read-access to the user's contacts.

In order to facilitate the machine learning detection approaches in this paper, we implemented a Java-based Android package analyzer and profiling tool for automated reverse engineering of the APK files. The steps involved are shown in Figure 1.

First, the .apk files are decompressed into separate folders containing the Manifest file, .dex file and other resource subfolders. Afterwards, the manifest file is converted into readable format using AXML2jar. The .dex file is then disassembled using a tool called Baksmali [24]. Baksmali is a disassembler for the dex format used by Dalvik. Baksmali disassembles .dex files into multiple files with .smali extensions. Each .smali file contains only one class information which is equivalent to a Java .class file. The files in the decompressed folders are mined to extract relevant properties subsequently used to construct the Bayesian classification-based models.

### 4. THE MACHINE LEARNING APPROACHES

Data mining and machine learning are increasingly being applied in the anti-malware industry, particularly in augmenting well-established heuristics and generics methods [25]. Data mining drives automation, which is motivated by reducing maintenance costs associated with the traditional heuristics and generics methods [25]. Data mining usually employs machine learning methods for inference, prediction, classification etc. Hence, it is important to select an appropriate method depending on the particular application. Bayesian classification is well suited to our problem of filtering large amounts of apps as it can perform relatively fast classification with low computational overhead once trained. Another important property which motivates its implementation in our approach for detecting suspicious Android applications, is the ability to model both an 'expert' and 'learning' system with relative ease compared to other machine learning techniques. Bayesian method allows the incorporation of prior probabilities (expert knowledge) even before the training phase. This hybrid property can be exploited as a performance tuning tool without incurring additional computational overhead.

### 4.1 The classifier model

The Bayesian based classifier consists of learning and detection stages. The learning stage uses a training set of known malicious samples in the wild and another set of benign Android applications, collectively called the app corpus. The Java-based package analyzer uses several 'detectors' to extract the desired features from each app in the corpus. The feature set is subsequently reduced by a feature ranking and selection function, while the training function calculates the marginal and conditional probabilities used in formulating the algorithm employed for the final classification decisions.

### 4.2 Feature ranking and selection

Let an application characteristic $r_i$ obtained from mining of the APKs by the analyzer, be defined by a random variable:

$$R_i = \begin{cases} 1, & \text{if discovered by the detectors} \\ 0, & \text{otherwise} \end{cases} \quad (1)$$

In order to ensure selection of the most relevant application features for the classification stage, we calculate the Mutual Information (MI) [26] or *information gain* of each feature $R_i$ with respect to the class variable C. This is used to rank the features and select the most relevant features during the feature selection stage prior to model training. Let C be a random variable representing the application class, *suspicious* or *benign*:

$$C \in \{suspicious, benign\}$$

Every application is assigned a vector defined by $\vec{r} = (r_1, r_2, ... r_n)$ with $r_i$ being the result of the $i^{th}$ random variable $R_i$. As the goal is to select the most relevant features, the feature selection function computes the MI ranking score of each random variable calculated as follows:

$$MI(R_i, C) = \sum_{r \in \{0,1\}} \sum_{c \in \{sus, ben\}} P(R_i=r; C=c) \cdot \log_2\left(\frac{P(R_i=r; C=c)}{P(R_i=r) \cdot P(C=c)}\right) \quad (2)$$

Given that $P(R_i=r; C=c) = P(R_i=r) \cdot P(C=c|R_i=r)$ the above equation becomes:

$$MI(R_i, C) = \sum_{r \in \{0,1\}} \sum_{c \in \{sus, ben\}} P(R_i=r) \cdot P(C=c|R_i=r) \cdot \log_2\left(\frac{P(C=c|R_i=r)}{P(C=c)}\right) \quad (3)$$

After calculating the score for each feature $R_i$, the feature set is then ranked in descending order and the top $n$ most relevant features with the highest information gain are then selected for training the model in order to maximize the classification accuracy.

### 4.3 Bayesian classification.

*A. Model evaluation*

According to Bayes theorem, the probability of an application with the feature vector $\vec{r} = (r_1, r_2, ... r_n)$ belonging in class C is given by:

$$P(C=c | \vec{R} = \vec{r}) = \frac{P(C=c) \prod_{i=1}^{n} P(R_i=r_i | C=c)}{\sum_{j \in \{0,1\}} P(C=c_j) \prod_{i=1}^{n} P(R_i=r_i | C=c_j)} \quad (4)$$

Where $P(R_i=r_i | C=c)$ and $P(C=c_j)$ are the estimated probabilities obtained from the frequencies calculated on the app learning corpus. While $n$ is the number of features used in the classification engine; $c_0$ and $c_1$ are the benign and suspicious classes respectively.

An app represented by the vector $\vec{r} = (r_1, r_2, ... r_n)$ is classified as benign if:
$$P(C=benign | \vec{R}=\vec{r}) > P(C=suspicious | \vec{R}=\vec{r}) \quad (5)$$

Otherwise, it is classified as suspicious. In terms of classification error, two cases can occur: (a) A benign app misclassified as suspicious. (b) A suspicious app misclassified as benign. In the context of our problem, the latter case is considered more critical, since allowing a malicious app to reach an end device is more critical than excluding a benign app from the distribution chain to be subject to further scrutiny.

### 4.4 Implemented Bayesian models from different data mining approaches

Three different data mining methods are implemented within the apk analyzer in order to build the Bayesian classification models. Through automated mining of the pre-processed .apk files, three separate models are built from:

- Input features derived from standard Android *permissions* extracted by static analysis of the Manifest files.
- Input features derived from *code-based properties* obtained by parsing disassembled .dex files present in the apk and other external resource files resulting from the apk decompression by the custom built analyzer.
- Input feature set consisting of a *mixture* of both standard permissions and code-based properties.

#### 4.4.1 *Permission-based Bayesian classifier*

Permissions are the most recognizable security feature in Android [22]. A user must accept them in order to install an application. Kirin [27] uses permissions for lightweight on-device application certification. Permissions have also been used in several of the Android tools mentioned in section 2, to provide app profiling information. Thus, their efficacy for machine learning based malware detection using trained models from large malware sample sets will be investigated. A permission is declared using the `<uses-permission>` tag in the Manifest file. For example, in order for an application to read phone contacts it must declare the standard Android permission as follows:

```
<uses-permission
android:name="android.permission.READ_CON
TACTS">
</uses-permission>
```

In order to build our permission-based model, 2000 Android applications comprising 1000 malware samples (from 49 different families) and 1000 benign apps were utilized. The apk analyzer parses the decrypted manifest file from each app and uses a *permissions detector* to match 131 standard Android permissions. Once a permission is detected, its count is incremented and stored. The stored total for each permission is further utilized by the feature selection function to rank and select the most relevant features for the permission-based Bayesian classifier, using equation (3).

The breakdown of the 49 malware families used and their respective number of samples are shown in Table 1. The malware samples were obtained from the Android Malware Genome Project [3]. The set of 1000 non-malicious apps were made up of different categories in order to cover a wide variety of application types. The categories include: entertainment, system tools, sports, health and fitness, news and magazines, finance, music and audio, business, education, games and a few other miscellaneous categories. The apps from third party market places were screened using *virustotal* scanning service to exclude potentially malicious apps from the benign set.

**TABLE 1. MALWARE FAMILIES USED AND THEIR NUMBERS.**

| FAMILY | NO OF SAMPLES | FAMILY | NO OF SAMPLES |
|---|---|---|---|
| ADRD | 22 | GINGERMASTER | 4 |
| ANSERVERBOT | 130 | GOLDDREAM | 47 |
| ASROOT | 8 | GONE60 | 9 |
| BASEBRIDGE | 100 | GPSSMSSPY | 6 |
| BEANBOT | 8 | HIPPOSMS | 4 |
| BGSERVE | 9 | JIFAKE | 1 |
| COINPIRATE | 1 | JSMSHIDER | 16 |
| CRUSEWIN | 2 | KMIN | 52 |
| DOGWARS | 1 | LOVETRAP | 1 |
| DROIDCOUPON | 1 | NICKYBOT | 1 |
| DROIDDELUXE | 1 | NICKYSPY | 2 |
| DROIDDREAM | 16 | PJAPPS | 58 |
| DROIDDREAMLIGHT | 46 | PLANKTON | 11 |
| DROIDKUNGFU1 | 30 | ROUGELEMON | 2 |
| DROIDKUNGFU2 | 34 | ROUGESPPUSH | 9 |
| DROIDKUNGFU3 | 144 | SMSREPLICATOR | 1 |
| DROIDKUNGFU4 | 80 | SNDAPPS | 10 |
| DROIDKUNGFUSAPP | 3 | SPITMO | 1 |
| DROIDKUNGFUUPDATE | 1 | TAPSNAKE | 2 |
| ENDOFDAY | 1 | WALKINWAT | 1 |
| FAKENETFLIX | 1 | YZHC | 22 |
| FAKEPLAYER | 6 | zHASH | 11 |
| GAMBLERSMS | 1 | ZITMO | 1 |
| GEINIMI | 69 | ZSONE | 12 |
| GGTRACKER | 1 | | |

The top 20 requested permissions extracted from the malware sample set are given in the Table 2. The top 20 permissions from the benign sample set are also shown in Table 3. Note that the top 20 permissions for malware samples were exactly as obtained in [3], whose Android Malware Genome project was the source of the malware samples used to build and analyze the models in our work1.

With the exception of ACCESS_NETWORK_STATE, INTERNET, WRITE_EXTERNAL_STORAGE and READ_PHONE_STATE, the top 10 requested standard permissions in our malware samples and benign set were different. It is interesting to note that READ_SMS, SEND_SMS, RECEIVE_SMS and WRITE SMS were amongst the 10 most requested in the malware samples but did not occur in the top 20 for our benign samples. (These were not in the top 20 of 1260 top free benign apps studied in [3] either, with the exception of SEND_SMS which was the 17th on the top 20 list in [3].)

**Table 2. Top 20 requested permissions from 1000 malware samples. The ranking corresponds to the findings in [3][1].**

| Permissions | Frequency |
|---|---|
| INTERNET | 939 |
| READ_PHONE_STATE | 888 |
| ACCESS_NETWORK_STATE | 741 |
| WRITE_EXTERNAL_STORAGE | 651 |
| READ_SMS | 591 |
| ACCESS_WIFI_STATE | 546 |
| RECEIVE_BOOT_COMPLETED | 497 |
| WRITE_SMS | 466 |
| SEND_SMS | 443 |
| RECEIVE_SMS | 394 |
| VIBRATE | 357 |
| ACCESS_COARSE_LOCATION | 355 |
| READ_CONTACTS | 344 |
| CALL_PHONE | 324 |
| ACCESS_FINE_LOCATION | 320 |
| WAKE_LOCK | 294 |
| WRITE_CONTACTS | 263 |
| CHANGE_WIFI_STATE | 251 |
| WRITE_APN_SETTINGS | 249 |
| RESTART_PACKAGES | 231 |

**Table 3. Top 20 requested permissions from 1000 benign samples.**

| Permissions | Frequency |
|---|---|
| INTERNET | 856 |
| ACCESS_NETWORK_STATE | 651 |
| WRITE_EXTERNAL_STORAGE | 471 |
| READ_PHONE_STATE | 388 |
| VIBRATE | 261 |
| ACCESS_COARSE_LOCATION | 245 |
| WAKE_LOCK | 234 |
| ACCESS_FINE_LOCATION | 221 |
| RECEIVE_BOOT_COMPLETED | 180 |
| ACCESS_WIFI_STATE | 176 |
| READ_CONTACTS | 102 |
| WRITE_SETTINGS | 93 |
| GET_ACCOUNTS | 88 |
| CAMERA | 85 |
| CALL_PHONE | 75 |
| WRITE_CONTACTS | 54 |
| GET_TASKS | 51 |
| RECORD_AUDIO | 51 |
| READ_HISTORY_BOOKMARKS | 41 |
| WRITE_HISTORY_BOOKMARKS | 35 |

This indicated that permissions attributes would provide discriminative capabilities for training the classifier to distinguish between malware and benign applications. In order to evaluate the permissions-based model, we carried out experiments designed to determine: (a) How effective the permissions-based features extracted from analysis of our malware and benign sample sets are in detecting unknown malware. (b) How well the permission-based model performs compared to the other viable models e.g. trained models derived from code properties extracted as features. Section 6 presents experimental results that provide some interesting insights.

---

[1] The top 20 permissions obtained from our benign set was also similar to the findings in [3], even though a different benign sample set of 1000 was used in this paper.

These are indeed pertinent questions given that a larger malware sample set covering more recent strains of Android malware is employed for our investigations compared to most previous works in Android malware detection that utilize machine learning. Also, permission based models provide a relatively lightweight static analysis approach since the need for reverse engineering of the .dex files and parsing a large number of files for feature extraction and classification is eliminated, resulting in considerable reduction of detection effort and time. Furthermore, permissions-based classification is useful because it is not susceptible to disassembly or decompilation failure which can sometimes hamper the reverse engineering during static analysis.

There are around 131 standard Android permissions that govern access to different system and device hardware resources. A user that intends to install an app will be prompted to accept or reject all the permissions requested by the app. In our model, we applied the analysis data for all of the 131 standard Android permissions to the feature selection function. The top ranked permissions (according to equation (3)) were subsequently selected for training the permissions-based Bayesian classifier. The top 30 ranked permissions and their respective information gain scores are shown in the Table 4.

**Table 4. Top 30 ranked permissions for the permission-based model (according to equation 3).**

| Ranked Permissions | Benign | Malware | Total | Infogain score |
|---|---|---|---|---|
| READ_SMS | 20 | 591 | 611 | 0.32920 |
| WRITE_SMS | 11 | 466 | 477 | 0.25053 |
| READ_PHONE_STATE | 388 | 888 | 1276 | 0.20962 |
| SEND_SMS | 24 | 443 | 467 | 0.20709 |
| RECEIVE_SMS | 14 | 394 | 408 | 0.19305 |
| WRITE_APN_SETTINGS | 4 | 249 | 253 | 0.12410 |
| ACCESS_WIFI_STATE | 176 | 546 | 722 | 0.11094 |
| RECEIVE_BOOT_COMPLETED | 180 | 497 | 677 | 0.08335 |
| INSTALL_PACKAGES | 10 | 199 | 209 | 0.08274 |
| CHANGE_WIFI_STATE | 31 | 251 | 282 | 0.08073 |
| CALL_PHONE | 75 | 324 | 399 | 0.07443 |
| RESTART_PACKAGES | 29 | 231 | 260 | 0.07289 |
| READ_CONTACTS | 102 | 344 | 446 | 0.06366 |
| WRITE_CONTACTS | 54 | 263 | 317 | 0.06351 |
| DISABLE_KEYGUARD | 21 | 155 | 176 | 0.04514 |
| READ_LOGS | 18 | 145 | 163 | 0.04382 |
| SET_WALLPAPER | 27 | 145 | 172 | 0.03482 |
| MOUNT_UNMOUNT_FILESYSTEMS | 14 | 115 | 129 | 0.03451 |
| READ_HISTORY_BOOKMARKS | 41 | 169 | 210 | 0.03351 |
| RECEIVE_WAP_PUSH | 1 | 60 | 61 | 0.02747 |
| WRITE_HISTORY_BOOKMARKS | 35 | 137 | 172 | 0.02537 |
| RECEIVE_MMS | 3 | 63 | 66 | 0.02487 |
| WRITE_EXTERNAL_STORAGE | 471 | 651 | 1122 | 0.02386 |
| READ_EXTERNAL_STORAGE | 19 | 99 | 118 | 0.02266 |
| GET_TASKS | 51 | 154 | 205 | 0.02168 |
| DELETE_PACKAGES | 7 | 61 | 68 | 0.01828 |
| CAMERA | 85 | 18 | 103 | 0.01793 |
| PROCESS_OUTGOING_CALLS | 10 | 66 | 76 | 0.01724 |
| ACCESS_LOCATION_EXTRA_COMMANDS | 33 | 103 | 136 | 0.01459 |
| INTERNET | 856 | 939 | 1795 | 0.01386 |

The impact of the ranking-based feature selection on near similar shared occurrences in permissions like ACCESS_NETWORK_STATE, ACCESS_COARSE_LOCATION, WAKE_LOCK, ACCESS_FINE_LOCATION, and VIBRATE can be clearly observed by their absence in Table 4, despite being in the top 20 permissions seen in both categories. It can also be observed with INTERNET permission being the 30[th] ranked feature.

The top ranked $n$ permissions were used to construct the input feature vectors $\vec{r} = (r_1, r_2, ... r_n)$ that characterize each application used in the training corpus. As mentioned earlier, $r_i$ is binary {0,1} indicating the presence or otherwise of the $i^{th}$ ranked permission in the feature vector.

*4.4.2 Code-based properties Bayesian classifier*

Unlike the permission-based model described above, the code-based model utilizes features extracted from code-based properties. A number of code-based properties were specified as matching criteria for a set of *property detectors* implemented within the apk analyzer. The detectors parse *.smali* files obtained from disassembled *.dex* files. In addition, external libraries, files within *assets* folders and *resources* folders are also scrutinized, if present within a decompressed APK.

The code-based properties matched by the detectors include: Android and Java API calls, Linux system commands, and some Android based commands and notifications. These provided a large feature set which were subsequently reduced to the top $n$ most relevant ones using the information gain criterion defined in equation 3. In total, we utilized 58 code-based properties for feature extraction. Our selection of these properties were guided by previous work (especially those that utilized similar properties for profiling Android apps and risk analysis) [2], [6], [16], as well as malware reports issued by mobile anti-virus vendors such as McAfee, and Lookout, detailing characteristics of malware discovered in the wild through manual analysis [32], [33]. Some of the described characteristics of several known malware families enabled us to define several corresponding matching properties for the property detectors that we employed for feature extraction. For example, concealment of secondary files in the resources or assets folders by sophisticated malware such as *Basebridge*, *Asroot* (which conceal shell scripts/commands to be executed at runtime), is the basis for defining features based on system commands such as 'chmod', 'mount', 'remount' 'chown', etc. The capabilities for dynamic code loading exhibited by families like *Plankton* also informed the choice of 'DexClassLoader' API calls and the inclusion of detecting embedded secondary '.*jar*' and '.*apk*' files as properties; while the use of encryption in malware such as *AnserverBot*, *Beanbot* etc., influenced the inclusion of cryptography API calls as property features.

In addition to attributes defined from domain knowledge gathered from the aforementioned sources, we included properties defined from observing outline profiles of hundreds of apps generated from our Java based APK analyzer and our lab-based study of publicly available malware samples from [30] and [31]. These profiles uncovered a high frequency of occurrences of some obvious properties (API calls) that indicated telephony services usage, Internet access, SMS activities, access to user contacts, messages and call logs etc., which facilitate theft of sensitive information and premium rate services access (both incentives for malware authors). Other additional properties we included in the feature set relate to the package manager API, presence of native code, the use of reflection-related API functions, and functions related to running background child processes.

In order to build the code-property based model, we applied the 58 properties to the same 2000 apps used for the permissions-based model. 10 out of these properties did not yield any match in the benign or malicious sample set, so were discarded. The remaining 48 were subsequently applied to the feature selection function which ranked them according to their scores. The top 25 ranked code-based properties and their respective frequencies in benign and malware categories are shown in Table 5.

**Table 5. Top 25 selected code-based properties and their frequencies in the benign and malware sets containing 1000 samples each (ranked using equation 3).**

| Properties | Benign | malware | Total | Infogain score |
|---|---|---|---|---|
| getSubscriberId (TelephonyManager) | 42 | 742 | 784 | 0.42853 |
| getDeviceId (TelephonyManager) | 316 | 854 | 1170 | 0.22919 |
| getSimSerialNumber (TelephonyManager) | 35 | 455 | 490 | 0.19674 |
| .apk (secondary payload) | 89 | 537 | 626 | 0.18202 |
| chmod (system command) | 19 | 389 | 408 | 0.17989 |
| abortBroadcast (intercepting broadcast notifications) | 4 | 328 | 332 | 0.17323 |
| intent.action.BOOT_COMPLETED | 69 | 482 | 551 | 0.16862 |
| Runtime.exec( ) (Executing process) | 62 | 458 | 520 | 0.16163 |
| /system/app | 4 | 292 | 296 | 0.15036 |
| getLine1Number (TelephonyManager) | 111 | 491 | 602 | 0.13116 |
| /system/bin | 45 | 368 | 413 | 0.12779 |
| createSubprocess (creating child process) | 0 | 169 | 169 | 0.08615 |
| remount (system command) | 3 | 122 | 125 | 0.05502 |
| DexClassLoader (stealthily loading a class) | 16 | 152 | 168 | 0.04953 |
| getSimOperator (TelephonyManager) | 37 | 196 | 233 | 0.04811 |
| pm install (installing additional packages) | 0 | 98 | 98 | 0.04725 |
| chown (system command) | 5 | 107 | 112 | 0.04325 |
| getCallState (TelephonyManager) | 10 | 119 | 129 | 0.04142 |
| /system/bin/sh | 4 | 90 | 94 | 0.03647 |
| .jar (secondary payload) | 87 | 252 | 339 | 0.03616 |
| mount (system command) | 29 | 152 | 181 | 0.03605 |
| KeySpec (code encryption) | 99 | 254 | 353 | 0.03067 |
| SMSReceiver | 3 | 66 | 69 | 0.02634 |
| getNetworkOperator (TelephonyManager) | 202 | 353 | 555 | 0.02071 |
| SecretKey (code encryption) | 119 | 248 | 367 | 0.02039 |

The table shows that some of the code-based properties such as 'pm install' and 'createSubprocess' were only found to be present in the malware sample set. References to system commands were also found mainly in the malware samples. References to .apk and .jar files which the detectors use to discover possible presence of secondary apps are found in both categories, but with more occurrences in the malware samples. Whilst secondary apps can be used to hide malicious payload, some legitimate apps such as popular ad and mobile payment frameworks are also know to utilize them [2]. As with the permissions-based model, the top ranked n code-based properties were used to construct the input feature vectors $\vec{r} = (r_1, r_2, ... r_n)$ that characterize each application used in the training corpus, after the feature selection stage.

### 4.4.3 Classifier based on combined ranked permissions and code-based properties

The third data mining approach that was implemented in the analyzer utilized a combination of permissions and code properties. The feature selection function was used to simultaneously rank the permissions and properties obtained from the code, using our 1000 benign and 1000 malware samples. The highest ranked from both were subsequently selected as input feature vectors for the Bayesian classifier model. The top 25 ranked from both permissions and code property-based feature selections are shown in Table 6. The top ten ranked had 5 permission-based and 5 code property-based properties. As can be seen from Table 6, the code properties were generally ranked higher within the top 25 than the permissions. This was because overall, more of the code property-based attributes had clearer discrepancies in their frequency in both categories than the permission based attributes. For this reason, code properties were likely to generate higher ranking scores than permissions.

**Table 6. Top 25 selected mixed features and their frequencies (ranked using equation 3).**

| Mixed Permission and code properties | Benign | malware | Total | Infogain score |
|---|---|---|---|---|
| getSubscriberId (TelephonyManager) | 42 | 742 | 784 | 0.42853 |
| READ_SMS | 20 | 591 | 611 | 0.32920 |
| WRITE_SMS | 11 | 466 | 477 | 0.25053 |
| getDeviceId (TelephonyManager) | 316 | 854 | 1170 | 0.22919 |
| READ_PHONE_STATE | 388 | 888 | 1276 | 0.20962 |
| SEND_SMS | 24 | 443 | 467 | 0.20709 |
| getSimSerialNumber (TelephonyManager) | 35 | 455 | 490 | 0.19674 |
| RECEIVE_SMS | 14 | 394 | 408 | 0.19305 |
| .apk | 89 | 537 | 626 | 0.18202 |
| chmod | 19 | 389 | 408 | 0.17989 |
| abortBroadcast | 4 | 328 | 332 | 0.17323 |
| intent.action.BOOT _COMPLETED | 69 | 482 | 551 | 0.16862 |
| Runtime.exec( ) | 62 | 458 | 520 | 0.16163 |
| /system/app | 4 | 292 | 296 | 0.15036 |
| getLine1Number (TelephonyManager) | 111 | 491 | 602 | 0.13116 |
| /system/bin | 45 | 368 | 413 | 0.12779 |
| WRITE_APN_SETTINGS | 4 | 249 | 253 | 0.12410 |
| ACCESS_WIFI_STATE | 176 | 546 | 722 | 0.11094 |
| createSubprocess | 0 | 169 | 169 | 0.08615 |
| RECEIVE_BOOT _COMPLETED | 180 | 497 | 677 | 0.08335 |
| INSTALL_PACKAGES | 10 | 199 | 209 | 0.08274 |
| CHANGE_WIFI_STATE | 31 | 251 | 282 | 0.08073 |
| CALL_PHONE | 75 | 324 | 399 | 0.07443 |
| RESTART_PACKAGES | 29 | 231 | 260 | 0.07289 |
| READ_CONTACTS | 102 | 344 | 446 | 0.06366 |

**4.5 Feature extraction times comparison**

The ranking and selection of top relevant features for training the models will significantly reduce computational overhead during the classification of applications, since the lower ranked 'redundant features' will not be utilized. This can be deduced from the time taken by our APK analyzer to extract the properties and construct feature vectors for training each of the models. In table 7, the average times taken to extract features from 516 reverse engineered apps using different feature settings are illustrated. The tests were performed on an Ubuntu 10.04 Linux PC running on 2.26 GHz Intel Xeon processor with 6GB of memory. When using top 25 mixed properties alone, the feature vectors were extracted from the 516 apps in 319 seconds (5 min 19s). In contrast, it took 1392 seconds (23 min 12s) for the analyzer to extract feature vectors consisting of all 58 code-based properties plus all the 131 permissions. Extracting the feature vectors for code-based properties alone took 1339 seconds (22 min 19s), while the vectors for the 131 permission based properties alone took 64 seconds to extract. Hence, at least 77% reduction in computational time can be achieved by feature reduction through the ranking and selection to reduce the entire feature space to the top 25 mixed features alone. The comparatively lower time taken to extract 131 permissions feature vectors for the 516 apps illustrates the characteristic of permissions based learning and classification as a relatively lightweight approach.

**Table 7. Feature vector extraction times from 516 apps for various attributes settings.**

| Attributes settings | Feature extraction time (s) |
|---|---|
| 25 top mixed attributes | 319 |
| 131 permissions only | 64 |
| 58 code properties only | 1339 |
| All 131 permissions and 58 code properties | 1392 |

## 5. METHODOLOGY AND EXPERIMENTS

As discussed earlier, our implementation of an APK analyzer includes the steps illustrated in Figure 1. The three models subsequently built were trained and tested under different feature selection settings in order to gain insight into their respective performances.

*5.1 Bayesian Classifier training*

For the training of the three Bayesian classification models, the same set of 2000 samples comprising 1000 malware and 1000 benign apps were used. In order to provide for testing and evaluation according to the evaluation criteria in equations (6) to (12) defined in the next sub-section, 5-fold cross validation was employed. Thus, 1600 samples (800 each of benign and malware) were used in the training, while the remaining 400 (200 each of benign and malware) were used for testing. Hence, the experiments undertaken used 5 different training and testing sets each containing a different testing portion with samples outside of its own training portion. This strategy was chosen to provide a wider range of samples for the testing of the classifiers' ability to detect *unknown malware*.

*5.2 Evaluation measures*

Several measures have been proposed in the literature for evaluating the predictive accuracy of machine learning based classifiers. These efficiency measures have been utilized in previous machine learning work [22], [28], [29], for example. In the context of our problem, the relevant measures utilized in our experiments are given below.

Let $n_{ben \rightarrow ben}$ be the number of benign applications correctly classified as benign, $n_{ben \rightarrow sus}$ the number of misclassified benign applications, $n_{sus \rightarrow sus}$ the number of suspicious applications correctly identified as suspicious while $n_{sus \rightarrow ben}$ represents the number of misclassified suspicious applications. Accuracy and Error Rate are respectively given by:

$$Acc = \frac{n_{ben \rightarrow ben} + n_{sus \rightarrow sus}}{n_{ben \rightarrow ben} + n_{ben \rightarrow sus} + n_{sus \rightarrow ben} + n_{sus \rightarrow sus}} \quad (6)$$

$$Err = \frac{n_{ben \rightarrow sus} + n_{sus \rightarrow ben}}{n_{ben \rightarrow ben} + n_{ben \rightarrow sus} + n_{sus \rightarrow ben} + n_{sus \rightarrow sus}} \quad (7)$$

The accuracy measurement indicates the overall proportion of correctly classified instances, whether suspicious or benign, during the testing phase of the particular model. The error rate given by (7) is the complementary measure to the accuracy, which can also be computed from Err =1-Acc. We also define *the false positive rate (FPR), false negative rate (FNR), true* *positive rate (TPR), true negative rate (TNR)* and *precision* (*p*) as follows:

$$FPR = \frac{n_{ben \to sus}}{n_{ben \to sus} + n_{ben \to ben}} \quad (8)$$

$$FNR = \frac{n_{sus \to ben}}{n_{sus \to sus} + n_{sus \to ben}} \quad (9)$$

$$TPR = \frac{n_{sus \to sus}}{n_{sus \to ben} + n_{sus \to sus}} \quad (10)$$

$$TNR = \frac{n_{ben \to ben}}{n_{ben \to sus} + n_{ben \to ben}} \quad (11)$$

$$p = \frac{n_{sus \to sus}}{n_{ben \to sus} + n_{sus \to sus}} \quad (12)$$

The false positive rate FPR, with respect to the suspicious class is measured by the proportion of misclassified true benign samples to the total number of benign sample instances during the testing phase. This is complementary to the true negative rate TNR, given by the proportion of the overall benign set that is correctly classified, illustrated by (11). Thus, true positive rate, TPR refers to truly malicious samples classified as suspicious divided by the overall number of malicious samples in the testing set. We also use the TPR and '*detection rate*' interchangeably, since this measure represents the model's capability to detect 'unknown' malicious samples. FNR measures the models tendency to misclassify suspicious apps as benign and is complementary to the detection rate. The precision reflects the precision of the model when it makes a decision to classify a sample as suspicious. Lastly, in our experiments, we also measured the AUC (Area under the Receiver Operator Characteristics (ROC) curve), i.e. the total area under the plot of TPR vs. FPR for every possible detection cut-off known as ROC. A perfect classifier will have an AUC of 1. Thus, the closer the AUC is to 1, the greater the model's predictive power.

6. RESULTS AND DISCUSSIONS

Figures 2 to 7 depict the results of experiments undertaken to evaluate the three implemented data mining approaches with Bayesian classifiers. The chart legends are suffixed with P, C and M to denote results from *Permission-based*, *Code property-based* and *Mixed* attributes respectively. Five different feature selection settings were used containing 5, 10, 15 and 20 features. Thus, 10f, 15f and 20f, represent the top 10, 15 and 20 ranked features according to the information gain from equation (3). 5fT refers to the 5 top features while 5fL refers to five lowest ranked from the top 20 (i.e. 16[th] to 20[th] ranked).

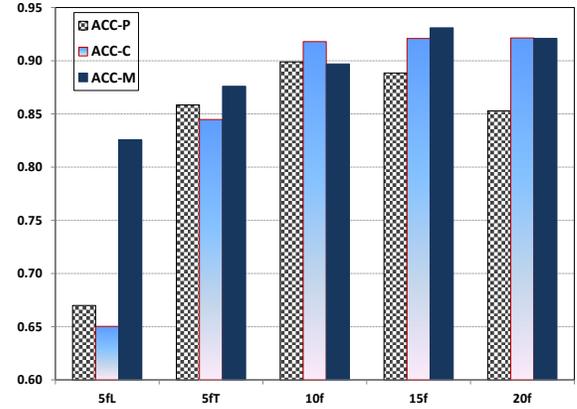

Figure 2: Average ACC for the three Bayesian models

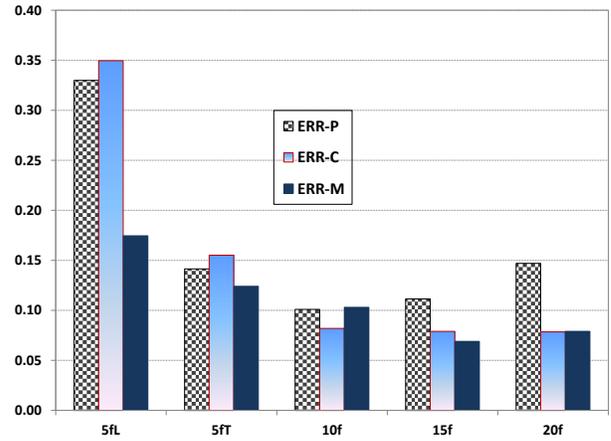

Figure 3: Average ERR for the three Bayesian models

From Figure 2, the results show that average accuracy improves with number of features selected for the C- and M-based models, while that of the P-based model peaks at 10 features. Correspondingly, Figure 3 depicts the average error rate decreasing for C- and M-based models as the features were increased, while P-based model recorded lowest error rate at 10 features. Overall, the best accuracy and error performance occurred when 15 features were used with the M-based Bayesian classifier. These are given in Table 9 as 0.931 and 0.069 respectively.

There was a large difference between the 5fL and 5fT results for P- and C based classifiers as seen in Figures 2 and 3. This highlights the effective selectivity of the feature selection function, since the same number of features but of different rankings were present in 5fL and 5fT feature sets. The 5fL features of the M-based model generally have a higher ranking than the 5fL features of both P- and C-based models; hence, its significantly better performance compared to the other two at the 5fL setting. As shown in table 8, the combined MI score for the 5fL features in the P-based model is 0.17413, while the 5fL features of the C-based model have a combined MI score of 0.2165. For the M-based model, the combined MI score of the 5fL features is 0.53172. This also accounts for the 5fL accuracy and error being relatively closer to that of 5fT for the M-based model when compared to the case with the P- and C-based models.

**Table 8. Information gain score comparison for the 5fL models.**

| Model | P-based 5fL | C-based 5fL | M-based 5fL |
|---|---|---|---|
| Combined Information gain score | 0.17413 | 0.21650 | 0.53172 |

Figures 2 and 3 also show that 15f accuracy/error performance is better than that of 20f for the M-based model. The plausible explanation for this can be found in Table 6. We notice that ACCESS_WIFI_STATE, RECEIVE_BOOT_COMPLETED, which form part of the 20f feature set, have a good number of occurrences in the benign category. The absence of these in the 15f set has the overall effect of reducing classification error rate. (This also accounts for the better TNR and FPR results of 15f than 20f for the M-based in Figures 6 and 7)

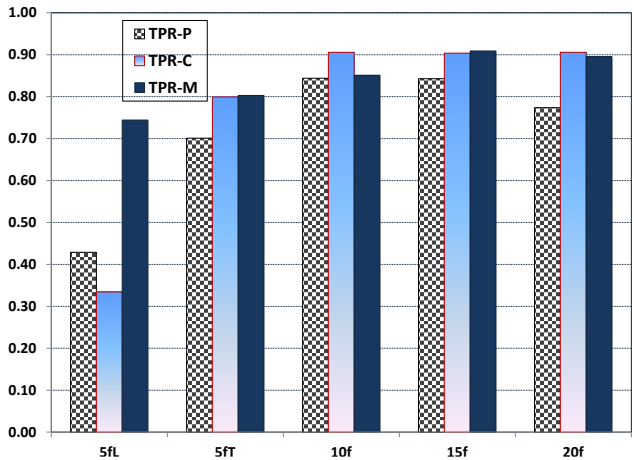

Figure 4: Average TPR for the three models.

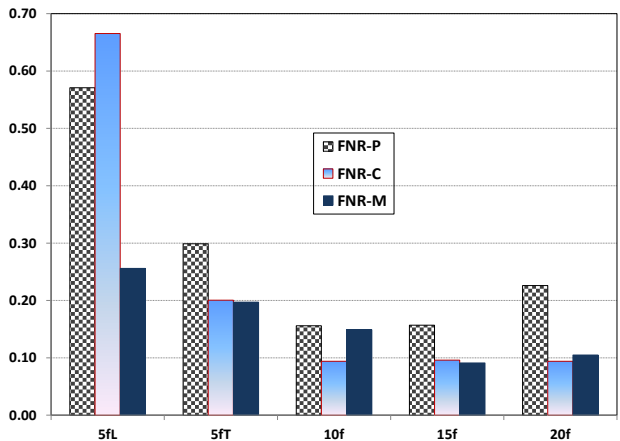

Figure 5: Average FNR for the three models.

Figure 4 depicts the TPR results for the three models with different feature settings. That is, the average rate of *unknown malware detection* by the trained models. The P-based model has lower detection rate than the M- and C- models at all feature settings, except at 5fL setting where C-based model is the lowest. The detection rates are quite similar for the 15f and 20f sets in C-based and M-based models. The actual values are shown in Table 9.

Overall, the best detection rate and hence lowest false negative rate were recorded with 15f used in the M-based model. In the context of our problem of filtering large app sample collections, a low false negative rate is highly desirable since this represents the proportion of 'missed' malware apps which may subsequently be installed as 'benign' apps. On that basis, the models based on M- or C-based features with the higher features settings should be preferred over the P-based model.

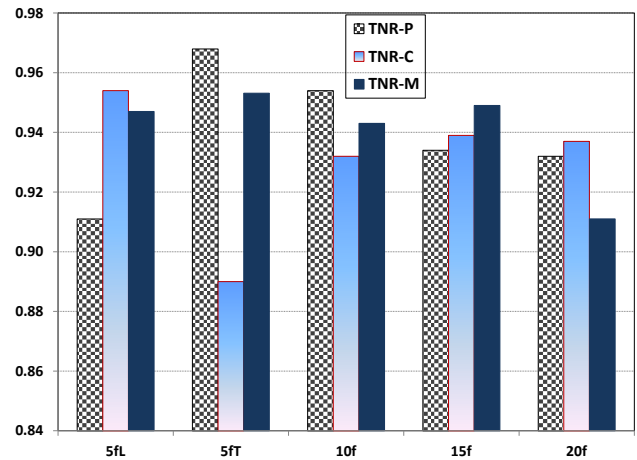

Figure 6: Average TNR for the three models.

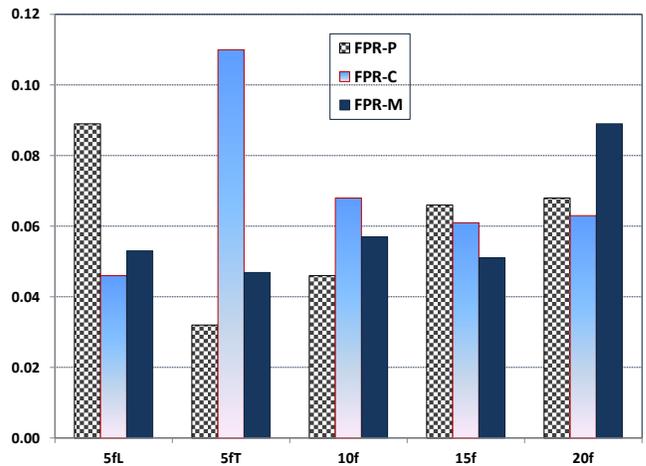

Figure 7: Average FPR for the three models.

In Figure 6, the average TNR results are illustrated. The exceptionally good performance shown by the 5fT for the P-based model can be attributed to 'sparse feature vectors' (as a result of fewer features used in model training) that will occur in a high proportion of benign samples (and also many of the malware samples). This leads to classifier bias towards benign class and hence high TNR; but, on the other hand it also results in higher FNR as can be seen clearly in Figure 5. Thus, we can conclude on that basis that M- or C-based models will still be preferable. Moreover, 5fT P-based model only yields about 70% detection rate.

Similarly, the exceptionally high FPR with 5fT for C-based model as seen in Figure 7 can be attributed to classifier bias due to 'sparse feature vectors' resulting from low number of features used for the model training. Again, from Table 9, 15f used with M-based model (which has the overall best accuracy/error performance) gave a reasonable low FPR of 0.051.

**Table 9. Summary of experimental results for the three models.**

|     | ERR-P | ERR-C | ERR-M | ACC-P | ACC-C | ACC-M |
| --- | --- | --- | --- | --- | --- | --- |
| 5fL | 0.330 | 0.350 | 0.175 | 0.670 | 0.650 | 0.826 |
| 5fT | 0.142 | 0.155 | 0.124 | 0.859 | 0.845 | 0.876 |
| 10f | 0.101 | 0.082 | 0.103 | 0.899 | 0.918 | 0.897 |
| 15f | 0.112 | 0.079 | **0.069** | 0.889 | 0.921 | **0.931** |
| 20f | 0.147 | 0.079 | 0.079 | 0.853 | 0.921 | 0.921 |

|     | TPR-P | TPR-C | TPR-M | FNR-P | FNR-C | FNR-M |
| --- | --- | --- | --- | --- | --- | --- |
| 5fL | 0.429 | 0.335 | 0.744 | 0.571 | 0.665 | 0.256 |
| 5fT | 0.701 | 0.799 | 0.803 | 0.299 | 0.201 | 0.197 |
| 10f | 0.844 | 0.906 | 0.851 | 0.156 | 0.094 | 0.149 |
| 15f | 0.843 | 0.904 | **0.909** | 0.157 | 0.096 | **0.091** |
| 20f | 0.774 | 0.906 | 0.895 | 0.226 | 0.094 | 0.105 |

|     | TNR-P | TNR-C | TNR-M | FPR-P | FPR-C | FPR-M |
| --- | --- | --- | --- | --- | --- | --- |
| 5fL | 0.911 | 0.954 | 0.947 | 0.089 | 0.046 | 0.053 |
| 5fT | **0.968** | 0.890 | 0.953 | **0.032** | 0.110 | 0.047 |
| 10f | 0.954 | 0.932 | 0.943 | 0.046 | 0.068 | 0.057 |
| 15f | 0.934 | 0.939 | 0.949 | 0.066 | 0.061 | **0.051** |
| 20f | 0.932 | 0.937 | 0.911 | 0.068 | 0.063 | 0.089 |

Table 10 shows the AUC (Area Under the ROC Curve) recorded for the three models at the various feature settings. An ROC curve plots the TPR against FPR for every possible detection cut-off. The total area under the ROC curve (AUC) indicates the classifier's predictive power. An AUC value of 1 implies perfect classification (i.e. 100% TPR and 0% FPR). Therefore, as mentioned earlier, an AUC value closer to 1 denotes better classifier predictive power. It can be observed from Table 9 that with the highest AUC of 0.97731, the M-based model with 15f setting is deemed the most predictive of all. Generally, lower AUC values were obtained by the P-based model compared to the C- and M-based models.

**Table 10. Area Under ROC curve and precision for all models.**

|     | AUC-P | AUC-C | AUC-M | Pre-P | Pre-C | Pre-M |
| --- | --- | --- | --- | --- | --- | --- |
| 5fL | 0.67103 | 0.61709 | 0.89217 | 0.825 | 0.860 | 0.894 |
| 5fT | 0.91377 | 0.94437 | 0.93859 | **0.960** | 0.880 | 0.940 |
| 10f | 0.93722 | 0.97428 | 0.96264 | 0.948 | 0.931 | 0.938 |
| 15f | 0.94259 | 0.97232 | **0.97731** | 0.927 | 0.937 | **0.950** |
| 20f | 0.94087 | 0.97223 | 0.97151 | 0.922 | 0.935 | 0.945 |

Precision results are also given for the three models in Table 10. Precision, as expressed in equation (11), denotes the precision of the model(s) when classifying samples as suspicious. It is therefore influenced by the number of false positives; a model with zero false positives will record 100% precision. From Table 9, it can be observed that the M-based model with 15f setting had precision of 0.950. Only 5fT with P-based model had a higher precision value. This, as mentioned earlier, can be attributed to classifier bias arising from 'sparse vectors' (due to the relatively small number of feature vectors) which enables relatively low false positive rate for the P-based model at 5fT setting as depicted in Figure 7.

The results suggest that mixed- based and code property-based models are a better choice than the permissions-only model. With overall accuracy values reaching approximately 0.9, 0.92, and 0.93 for the permission-based, code property-based, and mixed attributes models respectively, the three models recorded good performance. However, our comparative analyses with several metrics showed that mixed-based approach is the most promising of the three in the context of our problem, with potential for improvement. Note that the detection rates obtainable from all three models significantly exceed the best case of 79.6% with signature-based scanning recorded for the same malware sample set utilized in our experiments as reported in [3].

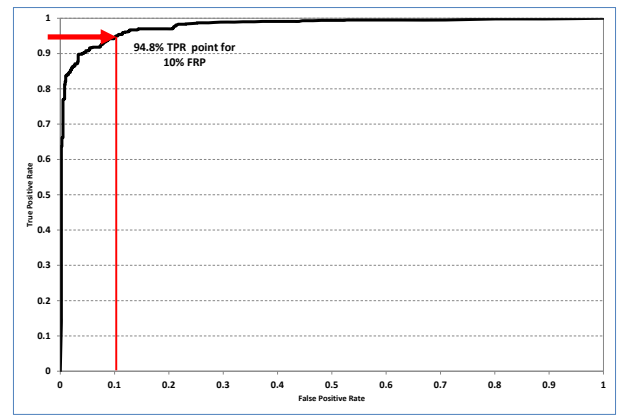

**Figure 8: The ROC curve for the 15f M-based model (AUC=0.97731)**

Another noteworthy aspect of our study is the excellent AUC performance of the best case model (i.e. 0.97731, from the 15f M-based model). The ROC plot is shown in Figure 8 below. The implication of a high AUC is that better detection rate performance can be obtained by trade-off with higher false positive rates which may be tolerable in some implementation scenarios. For instance, as part of an overall anti-malware system with further analysis stages, or in filtering apps to prioritize samples for further manual scrutiny, or as an input stage to drive decisions such as length and depth of further analysis processes. Figure 8 depicts the model's ability to operate at 94.8% detection rate with 10% FPR.

The results of this paper compare favourably with related works in the literature, thus highlighting the significance of our approach. Previous related work which employed static analysis used different sample sizes than ours, so a direct comparison is not straightforward (although some of the malware samples across these studies overlap). For instance, [29] employed 238 malware samples while [22], [34] and [35] based their experiments on 249, 378 and 121 malware samples respectively. Our study, on the other hand, utilized 1000 malware samples but nevertheless performed competitively and for most performance metrics outperformed the previous models. For instance, our 15f M-based model had a detection rate of 0.91 compared to 0.873 in [29] and was close enough to the best case of 0.92 in [22] despite using a much larger sample set. On the other hand the AUC of our model is significantly higher (0.97731) compared to 0.92 best case obtained in [22], which accounts for their false positive rate of 0.21 being much higher than the false positive rate of 0.051 obtained with our 15f M-based model. In Figure 9 we compare AUC results from this paper with previous work, highlighting the excellent predictive power of our (M-based model) approach. The AUC from our 15f C-based model was 0.972, also higher than previously published results. Our 15f P-based model, which was the best case for the permissions only scenario, also performed very well with an AUC of 0.9426.

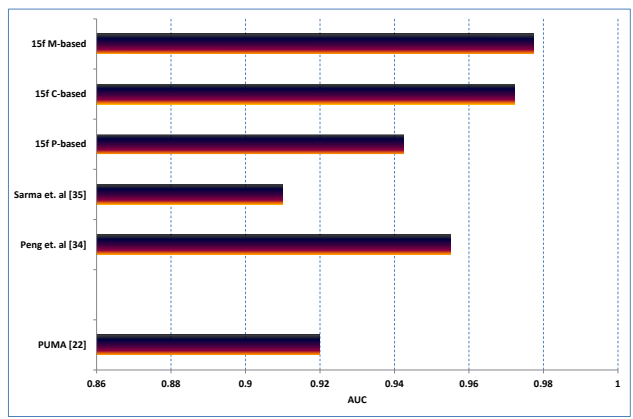

**Figure 9: AUC results comparison with related work.**

## 7. CONCLUSION

In this paper, we investigated three data mining based methods for detecting unknown Android malware. These utilized Bayesian classification models built from mining data generated by automated reverse engineering of the Android application packages using a Java implemented custom package analyzer. The three models investigated were built from static analysis of (a) Standard Android permissions in the Manifest files (b) Code properties indicative of potential malicious payload (c) both standard permissions and code properties. The models were built by extracting these properties from a set of 1000 samples of 49 Android malware families together with another 1000 benign applications across a wide variety of categories.

Extensive experiments were undertaken to study the performance of the models in terms of error rate, accuracy, true negative rate, true positive rate, false positive rate, false negative rate, precision and also area under ROC curve. The results suggest that mixed- based and code property-based models are a better choice than the permissions-only model. With overall accuracy values reaching approximately 0.9, 0.92, and 0.93 for the permission-based, code property-based, and mixed attributes models respectively, the three models recorded good performance. However, our comparative analyses with several metrics showed that mixed-based approach is the most promising of the three in the context of our problem. With this method, an excellent predictive power evidenced by AUC result of about 0.977 is achievable, exceeding previous similar approaches in the published literature.

Our results not only demonstrate practically the potential of data mining for unknown Android malware detection, but also the effectiveness of the Bayesian classification models for tackling this problem. Thus, the models provide a complementary approach to signature-based scanning or dynamic analysis, and fast filtering capabilities for large scale analyses to uncover unknown malware. The malware samples used in our experiments were from the largest publicly available collection at the time of writing. Hence, future work would investigate the models' performance with larger sample sets as more malware samples are discovered in the wild.